\documentclass[]{article}

\begin{document}
\newcommand{\bgm}[1]{\mbox{\boldmath $#1$}}
\newcommand{\bgms}[1]{\mbox{{\scriptsize \boldmath $#1$}}}
\newcommand{\ul}[1]{\underline{#1}}
\newcommand{\bgt}[1]{{\boldmath $#1$}}
\hyphenation{Bia-lek Eg-el-haaf Ruy-ter War-ze-cha}

\leftline{\Large\bf Real time encoding of motion:}
\leftline{\Large\bf Answerable questions and questionable answers}
\leftline{\Large\bf from the fly's  visual system}
\bigskip\bigskip

\leftline{\large  Rob de Ruyter van Steveninck,$^1$ Alexander
Borst,$^{2}$ and William Bialek$^{1}$}
\bigskip

\leftline{$^1$NEC Research Institute}
\leftline{4 Independence Way}
\leftline{Princeton, New Jersey 08540 USA}
\leftline{$^2$ESPM---Division of Insect Biology}
\leftline{ University of California at
Berkeley}
\leftline{201 Wellman Hall} 
\leftline{Berkeley,  California 94720, USA}
\bigskip


\bigskip\bigskip\hrule\bigskip\bigskip

\noindent In the past decade, a small corner of the fly's
visual system has become an important testing ground for
ideas about coding and computation in the nervous system.
A number of results demonstrate that this system operates
with a precision and efficiency near the limits imposed by physics, and
more generally these results point to the reliability and
efficiency of the strategies that nature has selected for
representing and processing visual signals.  A recent series of
papers by Egelhaaf and coworkers, however,   suggests that almost all
these conclusions are incorrect.  In this contribution we
place these controversies in a larger context, emphasizing
that the arguments are not just about flies, but rather about how we
should quantify the neural response to complex, naturalistic
inputs.  As an example,  Egelhaaf et al. (and many others) compute
certain correlation functions and use the apparent correlation times as a
measure of temporal precision in the neural response.  This
analysis neglects the structure of  the correlation function at short
times, and we show how to analyze this structure to reveal a temporal
precision 30 times better than suggested by the correlation time; this 
precision is confirmed by a much more detailed information theoretic
analysis.  In reviewing other aspects of the controversy, we find that
the analysis methods used by Egelhaaf et al. suffer from some
mathematical inconsistencies, and that in some cases we are unable to
reproduce their experimental results. Finally, we  present results from
new experiments that probe the neural response to inputs that approach
more closely the natural context for freely flying flies. These new
experiments  demonstrate that the fly's visual system is even more
precise and efficient under natural conditions than had been inferred
from our earlier work. 

\vfill\newpage

\section{Introduction}

Much of what we know about the neural processing of sensory information
has been learned  by studying the responses of single neurons to rather
simplified stimuli. The ethologists,  however, have argued that we can
reveal the full richness of the nervous system only when  we study the
way in which the brain deals with the more complex stimuli that occur in
nature.  On the other hand it is possible that the processing of natural
signals is decomposable into  steps that can be understood from the
analysis of simpler signals. But even then, to prove that  this is the
case one must do the experiment and use complex natural stimuli. In the
past  decade there has been renewed interest in moving beyond the simple
sensory inputs that have  been the workhorse of neurophysiology, and a
key step in this program has been the  development of more powerful tools
for the analysis of neural responses to complex dynamic  inputs. The
motion sensitive neurons of the fly visual system have been an important
testing  ground for these ideas, and there have been several key results
from this work: 
\begin{enumerate}
\item The sequence of spikes from a motion sensitive neuron
can be decoded to recover a  continuous estimate of the dynamic velocity
trajectory (Bialek et al. 1991; Haag and Borst  1997). In this decoding,
individual spikes contribute significantly to the estimate of  velocity
at each point in time.
\item The precision of velocity estimates approaches
the physical limits imposed by diffraction  and noise in the
photoreceptor array (Bialek et al. 1991). 
\item One or two spikes are
sufficient to discriminate between motions which differ by  displacements
in the `hyperacuity' range, an order of magnitude smaller than the
spacing  between photoreceptors in the retina (de Ruyter van Steveninck
and Bialek 1995). Again  this performance approaches the limits set by
diffraction and receptor noise.
\item Patterns of spikes which differ by
millisecond shifts of the individual spikes can stand for 
distinguishable velocity waveforms (de Ruyter van Steveninck and Bialek
1988), and  these patterns can carry much more information than expected
by adding up the  contributions of individual spikes (de Ruyter van
Steveninck and Bialek 1988, Brenner et  al. in press).
\item The total
information that we (or the fly) can extract from the spike train
continues to  increase as we observe the spikes with greater temporal
resolution, down to millisecond  precision (de Ruyter van Steveninck et
al. 1997, Strong et al. 1998). 
\item These facts about the encoding of
naturalistic, dynamic stimuli cannot be extrapolated  simply from studies
of the neural response to simpler signals. The system exhibits  profound
adaptation (Maddess and Laughlin 1985, de Ruyter van Steveninck et al.
1986,  Borst and Egelhaaf 1987, de Ruyter van Steveninck et al. 1996,
Brenner et al. submitted),  so that the encoding of signals depends
strongly on context, and the statistical structure of  responses to
dynamic stimuli can be very different from that found with simpler static
or  steady state stimuli (de Ruyter van Steveninck et al. 1997). 
\end{enumerate}
We emphasize that many of these results from the fly's visual system have
direct analogs in  other systems, from insects to amphibians to primates
(Rieke et al. 1997). 

In a series of recent papers, Egelhaaf and coworkers
have called these results into question  (Warzecha and Egelhaaf 1997,
1998, 1999; Warzecha et al. 1998). Several of these papers are  built
around a choice of a stimulus very different from that used in previous
work. Rather than  synthesize a stimulus with known statistical
properties, they sample the time dependent  motion signals generated by a
fly tethered in a flight simulator. The simulator is operated in  closed
loop so that the fly, by producing a yaw torque which is measured
electronically,  moves a pattern on a CRT monitor, while the animal
itself stays stationary. For experiments  on the responses of the motion
sensitive neurons these patterns and motions are replayed to  another
fly, again through a monitor. In their judgement these 
stimuli ``are characteristic of a  normal behavioral 
situation in which the actions and reactions of the 
animal directly affect its  visual input'' (Warzecha 
and Egelhaaf 1998).

For these stimuli, Warzecha and Egelhaaf claim that the
timing of individual spikes has no  significance in representing motion
signals in the fly's motion sensitive neurons. Instead they  suggest that
the neuron's response should be averaged over time scales of order 40 to
100 ms  to recover the essential information, and that timing of spikes
within this averaging window is  irrelevant. These claims are in conflict
with points [1], [4], and [5] above. As part of their  discussion of
these points Warzecha and Egelhaaf make repeated references to the
noisiness  of the neural response, in apparent contradiction of points
[2] and [3], although they do not  address specifically the quantitative
results of the earlier work. Finally, they suggest that there  is no
difference between the statistics of spike trains in response to steady
state vs. dynamic  stimuli, in contradiction of point [6]. 

Obviously the
recent work of Egelhaaf and coworkers raises many different issues. In
this  contribution we try to focus on three problems of general interest.
First, how do we define a  meaningful ``naturalistic stimulus,'' and does
their ``behaviourally generated'' stimulus fall into  this category? In
particular, how do we reach an effective compromise between stimuli that 
occur in nature and stimuli that we can control and reproduce reliably in
the laboratory?  Second, how do we characterize the neural response to
complex dynamic inputs? In  particular, how do we evaluate all the
relevant time scales in the sensory signal itself and in  the spike
train? Again, these are issues that we must face in the analysis of any
neural system  for processing of sensory information; indeed there are
even analogous issues in motor  systems. Thus the fly's visual system
serves here as an example, rather than as an end in itself. 

Before we
begin our discussion of these two points, we must be clear that the first
question---what is a natural stimulus?---is a question about the biology
and ecology of the animal we are  studying, as well as a question about
the design and constraints of a particular experimental  setup. One might
well disagree about the best strategy for generating naturalistic stimuli
in the  lab. On the other hand, our second question---how do we
characterize the response to  complex signals?---is a theoretical issue
which is not tied to the particulars of biology. On this  issue there are
precise mathematical statements to be made, and we hope to make clear
how  these mathematical results can be used as a rigorous guide to the
analysis of experiments. 

The third and final question we address concerns
the comparison between static and dynamic  stimuli. Although we believe
that the most interesting problems concern the way in which the  brain
deals with the complex, dynamic stimuli that occur in nature, much has
been learned  from simpler static stimuli and there are nagging questions
about whether it really is  `necessary' to design new experiments that
need more sophisticated methods of analysis. For  reasons that will
become clear below, the comparison of static and dynamic stimuli also is 
crucial for understanding whether many of the lessons learnt from the
analysis of the fly's  motion sensitive neurons will be applicable to
other systems, especially the mammalian  cortex. 

\section{What is a natural stimulus?}

The fly's motion sensitive neuron H1 offers a relatively simple
testing ground for ideas about  the neural representation of natural
signals. This cell is a wide field neuron, so rather than  coding the
motion of small objects or a component of the local velocity flow field,
H1 is  responsible primarily for coding the rigid body horizontal (yaw)
motion of the fly relative to  the rest of the world. Thus there is a
limit in which we can think of ``the stimulus'' as being a  single
function of time, $\nu(t)$, which describes this angular velocity
trajectory. It should be  clear that this description is incomplete: the
neural response is affected also by the mean light  intensity, the spatial
structure of the visual stimulus, and the area of the compound eye that
is  stimulated. Further, the system is highly adaptive, so that the
encoding of a short segment of  the trajectory  $\nu(t)$ will depend
strongly on the statistics of this trajectory over the past several 
seconds. 

Traditional
experiments on motion sensitive neurons (as on other sensory cells) have
used  constant stimuli (motion at fixed velocity), pulsed stimuli
(stepwise motion), or have analysed  the steady state behaviour in
response to sinusoidal motion at different frequencies. In nature, 
trajectories are not so simple. Instead one can think of trajectories as
being drawn from a  distribution $P[ \nu(t)]$ or ``stimulus ensemble.'' A
widely used example of stimulus ensembles is  the Gaussian ensemble, in
which the distribution of trajectories is described completely by the 
spectrum or correlation function. We can construct spectra and correlation
functions so that  there is a single characteristic stimulus
amplitude---the dynamic range $\nu_{\rm rms}$ of velocity  signals---and a
single characteristic time $\tau_c$ in the dynamics of these signals. A
reasonable  approach to the study of naturalistic stimuli might then be
to explore the coding of signals in  H1 using stimulus ensembles
parametrized by $\nu_{\rm rms}$ and $\tau_c$. Most of the results
enumerated  above have been obtained in this way. 

In their recent papers (Warzecha and Egelhaaf 1997, Warzecha
et al. 1998), as well as in their  contribution to this volume, Warzecha
and Egelhaaf argue that the stimulus ensembles used in  experiments on H1
have been restricted unfairly to short correlation times. Put another
way,  the stimuli used in these experiments have included high temporal
frequency components.  Warzecha and Egelhaaf suggest that these high
frequency components bias the response of the  motion sensitive cells to
artificially high temporal precision which is not relevant for the 
behaviourally generated stimuli that they use.\footnote{In fact Warzecha and Egelhaaf make two different arguments about
high frequency stimuli. They make  repeated references to the integration
times and noise in the fly's visual system, all of which limit the
reliability  of responses to high frequency components in the input.
These arguments generally are presented in qualitative  terms, but
Warzecha and Egelhaaf (1999)  state explicitly that signals above 30 Hz
are undetectable above the  noise and hence can have no impact on the
statistics of the spike train. On the other hand, Warzecha and  Egelhaaf
(1997) argue that the inclusion of high frequency components in the input
causes an unnaturally tight  locking of spikes to stimulus events,
causing us to overestimate the significance of spike timing for the
coding of  behaviorally relevant stimuli. It should be clear that these
two arguments cannot both be correct.}
The question of whether
timing precision is  important under truly natural conditions is left
open.  

Independent of what is truly natural, one can argue that
experiments with short correlation  times have provided evidence on what
the fly's visual system can do. Although we seldom sit  in dark rooms and
wait for dim flashes of light, such experiments led to the demonstration 
that the human visual system can count single photons (Hecht et al.
1942). In this spirit,  studies of H1 using stimuli with short
correlation times have revealed that the fly's nervous  system can
estimate velocity with a precision limited by noise in the photoreceptor
array and  that timing relations between neural responses and stimulus
events can be preserved with  millisecond precision, even as the signals
pass through four stages of neural circuitry. It  would seem strange that
such impressive performance would evolve if it were irrelevant for  fly
behaviour. 

Instead of choosing trajectories $\nu(t)$ from a known probability
distribution, we could try to  sample the trajectories that actually
occur in nature. Here we have to make choices, and these  will always be
somewhat subjective: Dethier (1976) reports that female flies spend
12.7\%,  and male flies 24.3\% of their time walking or flying. The other
activities on Dethier's list are  feeding, regurgitating, grooming and
resting, during which information from the fly's motion  sensitive cells
presumably is not too relevant. So it seems the fly could live quite
happily  without its tangential cells most of its time. On the other
hand, during periods of flight, the  responses of its motion sensitive
cells are strongly modulated. On top of that, the depth and  speed of
modulation may vary as the fly switches from periods of relatively quiet
cruising to  episodes of fast and acrobatic pursuit or escape, and back
(Land and Collett 1974). Although  it is not clear at the outset what
portion of the total behavioural repertoire we should analyse,  the thing
that presumably tells us most about the ``design'' of the fly is the
dynamics of neural  signal processing during top performance.
Correspondingly, Warzecha and Egelhaaf propose  to use stimuli that are
representative of the trajectories experienced by a fly in flight, and
we  agree that this is an excellent choice. There are still some
difficulties, however. Warzecha and  Egelhaaf propose that meaningful
data can be obtained from ``behaviourally generated''  trajectories
$\nu(t)$ recorded from flies that are tethered in a flight simulator
apparatus in which  the fly's measured torque is to 
move a pattern on a CRT monitor in the visual field of 
the fly.  The combination of fly, torque meter, and 
moving pattern thus acts as a closed loop feedback 
system whose dynamical properties are determined both by
the fly and by the gain and  bandwidth of the mechanical
and electronic components involved. The data presented 
by  Warzecha and Egelhaaf (1997, 1998, and this volume) 
strongly suggest that the dynamics of  the feedback 
system are dominated by the electromechanical properties
of their setup, and not  by the fly itself. This is most
clearly seen from direct comparisons between the 
trajectories in  the flight simulator and those observed
in nature.

Trajectories during free flight were recorded in the classic work of Land
and Collett (1974),  who studied chasing behaviour in Fannia canicularis
and found turning speeds of several  thousand degrees per second.
Wehrhahn (1979), Wehrhahn et al. (1982) and Wagner  (1986a,b,c) report
very similar results for the housefly Musca, and recent publications 
(Schilstra and van Hateren 1998, van Hateren and Schilstra 1999) report
flight measurements  at high temporal and spatial resolution, from
Calliphora flying almost free. In their published  dataset flies made
about 10 turns per second, during which head velocities easily exceeded 
1000$^\circ$/s, while maximum head turning velocities were well over
3000$^\circ$/s. If we compare the  results of these studies to the motion
traces used in the experiments by Warzecha and  Egelhaaf (1997, 1998) we
see that their traces are considerably smoother, and do not go  beyond
100$^\circ$/s. These differences are illustrated in Fig. 1, where we make
an explicit  comparison between free flight data obtained by Land and
Collett (1974) and the motion  traces data presented in Fig. 1 of
Warzecha and Egelhaaf (1997). It is clear that there are  dramatic
differences in the frequency of alternation and, especially, in the
amplitude of the  motion signals. We are not sure how Egelhaaf and
Warzecha can maintain their claim that  ``there are likely to be few
instances in the normal world where visual motion encompasses a  wider
dynamic range than that which could be tested here'' (Warzecha et al.
1998, p. 362). Simple theoretical arguments suggest that these
differences between the flight simulator  trajectories and true natural
trajectories will have enormous consequences for the reliability of 
responses in the motion sensitive neurons. Warzecha and Egelhaaf (Fig 6
of their keynote  paper in this volume) report estimates of the signal
and noise power spectra in the graded  voltage response of a motion
sensitive cell. If we scale the signal to noise power ratio they  present
in proportion to the ratio between the power spectrum of natural motion
and the  velocity power spectrum they used, then the signal to noise
ratio will increase so much that  the natural trajectories will produce
signal resolvable against the noise at frequencies well  above 200 Hz.
This would mean that events in natural stimuli will be localizable with 
millisecond precision.  

There are other differences between the stimulus conditions
studied by Warzecha and Egelhaaf and the natural conditions of 
free flight. Outdoors, in the middle of the afternoon, light
intensities typically are two orders of magnitude larger than are
generated with standard laboratory displays (Land 1981). Further,
the wide field motion sensitive cells gather inputs from large
portions of the compound eye (Gauck and Borst 1999), which
extends backward around the head to cover a large fraction of the
available solid angle; rotation of the fly produces coherent
signals across this whole area, and it is very difficult to
reproduce this ``full vision'' in the lab with CRT displays.
While it is difficult to predict quantitatively the consequences
of these differences, the qualitative effect is clear: natural
signals are much more powerful and ``cleaner'' than the stimuli
which Warzecha and Egelhaaf have used.

We can take a
substantial step toward natural stimulus conditions by recording from a
fly that  itself rotates in a natural environment along a trajectory
representative of free flight.  Preliminary results from such experiments
will be analysed in more detail below, and a  detailed account is
forthcoming (Lewen et al. in preparation). A female wild fly
(Calliphora),  caught outdoors, was placed in a plastic tube and
immobilized with wax. A small incision was  made in the back of the head,
through which a microelectrode could be advanced to the lobula  plate to
record from H1. The fly holder, electrode holder and manipulator were
assembled to  be as light and compact, yet rigid, as possible. In this
way the fly and the recording setup  could be mounted on the axle of a
stepper motor (Berger-Lahr, RDM 564/50, driven by a  Divi-Step D331.1
interface with 10,000 steps/revolution) and rotated at speeds of up to 
several thousand degrees per second. The motor speed was controlled
through the parallel  port of a laptop computer by means of custom
designed electronics, and was played out at  2ms intervals. The data
presented here are from an experiment in which the setup was placed 
outside on a sunny day, in a wooded environment not far from where the
fly was caught. A  simple, but crucial, control is necessary: H1 does not
respond if the fly is rotated in the dark,  or if the visual scene
surrounding it rotates together with the fly. We can thus be confident 
that H1 is stimulated by visual input alone, and not by other sensory
modalities, and also that   electronic crosstalk between the motor and
the neural recording is negligible. The motion trace $\nu(t)$ was derived
from a concatenation of body angle readings over the  course of the
flight paths of a leading and a chasing fly as depicted in Fig. 4 of Land
and  Collett (1974). For technical reasons we had to limit the velocity
values to half those derived  from that figure, but we have no reason to
believe that this will affect the main result very  much. Translational
motion components were not present, representing a situation with 
objects only at infinity. Padded with a few zero velocity samples, this
trace was 2.5 seconds  long. That sequence was repeated with the sign of
all velocity values changed, to get a full 5  second long sequence. This
full sequence was played 200 times in succession while spikes  from the
axon terminals of H1 were recorded as an analog waveform at 10 kHz
sampling rate.  In off line analysis spike occurrence times were derived
by matched filtering and  thresholding.  

Before looking at the responses
of H1, we emphasize several aspects of the stimulus  conditions: 
\begin{itemize}
\item The
motion stimulus is obtained from direct measurement of flies in free
flight, not from a  torque measurement of a tethered fly watching a CRT
monitor. As argued above, the  electromechanical properties of the setup
used by Warzecha and Egelhaaf are likely to  have drastic effects on the
frequency and amplitude characteristics of the motion. 
\item The field of
view experienced by the fly in our setup is almost as large as that for a
free  flying fly. Most of the visual field is exposed to movement, with
the exception of a few  elements (e.g. the preamplifier) that rotate with
the fly, and occupy just a small portion of  the visual field.
\item The
experiment is done outside, in an environment close to where our
experimental flies  are caught, so that almost by definition we stimulate
the fly with natural scenes. 
\item The experiment is performed in the afternoon on a bright day. 
>From dim to bright patches  of the visual scene the effective estimated
photon flux for fly photoreceptors under these  conditions varies from
$5\times 10^5$ to $5\times 10^6$ photons per second per receptor. Warzecha
and  Egelhaaf's experiments (as many experiments of ours) were done with
a fly watching a  Tektronix 608 cathode ray tube, which has an estimated
maximum photon flux of about  $10^5$ photons per second per receptor. 
\end{itemize}
Figure 2 shows the spike trains generated by H1 in the 
``outdoor'' experiment, focusing on a  short segment of 
the experiment just to illustrate some qualitative 
points. The top trace shows  the velocity waveform 
$\nu(t)$, and subsequent panels show the spikes 
generated by H1 in  response to this trajectory (H1$+$) 
or its sign reverse (H1$-$). Visual inspection reveals 
that some  aspects of the response are very 
reproducible, and further that particular events in the 
stimulus  can be associated reliably with small numbers 
of spikes. The first stimulus zero crossing at  about 
1730 ms is marked by a rather sharp drop in the activity
of H1$+$, with a sharp rise for  H1$-$. This sharp 
switching of spike activity is not just a feature of 
this particular zero  crossing, but occurs in other 
instances as well. Further, the small hump in velocity 
at about  2080 ms lasts only about 10 ms, but induces a 
reliable spike pair in H1$+$ together with a short  
pause in the activity of H1$-$. The first spike in H1$-$
after this pause (Fig. 2c) is timed quite  well; its 
probability distribution (Fig. 2e) has a standard 
deviation of 0.73 ms. Thus, under  natural stimulus 
conditions individual spikes can be locked to the 
stimulus with millisecond  precision.

In fact the first few
spikes after the pause in H1$-$ have even greater internal or relative 
temporal precision. The raster in Fig. 2c shows that the first spike
meanders, in the sense that  the fluctuation in timing from trial to
trial seems to be slow. This suggests that much of the  uncertainty in
the timing of this spike is due to a rather slow process, perhaps
metabolic drift.  To outside observers, like us, these fluctuations just
add to the spike timing uncertainty,  which even then is still
submillisecond. Note, however, that to some extent the fly may be  able
to compensate for that drift. If the effect is metabolic, then different
neurons might drift  more or less together, and the time interval between
spikes from different cells could be  preserved quite well in spite of
temporal drift of individual spikes. Similarly, within one cell,  spikes
could drift together (Brenner et al., 2000), and this indeed is the case
here. As a result  the interval between the first spike and the next is
much more precise, with a 0.18 ms standard  deviaton, and it does not
seem to suffer from these slow fluctuations (Fig. 2d). The timing 
accuracy of ensuing intervals from the first spike to the third and
fourth, although becoming  gradually less well defined, is still
submillisecond (Fig. 2f). So it is clear that some identifiable patterns
of spikes are generated with a timing precision of the order of a 
millisecond or even quite a bit better. 

Although we have emphasized the
reproducibility of the responses to natural stimuli, there  also is a
more qualitative point to be made. All attempts to characterize the
input/output  relation of H1 under laboratory conditions have indicated
that the maximum spike rate should  occur in response to velocities below
about 100$^\circ$/s, far below the typical velocities used in our 
experiments. Indeed, many such experiments suggest that H1 should shut
down and not spike  at all in response to these extremely high
velocities. In particular, Warzecha and Egelhaaf  (1998) claim that spike
rates in H1 are essentially zero above 250$^\circ$/s, that this lack of 
sensitivity to high speeds is an essential result of the computational
strategy used by the fly in  computing motion, and further that this
behaviour can be used to advantage in optomotor  course control. The
outdoor experiment demonstrates that none of these conclusions are 
relevant to more natural conditions, where H1's response peaks at about
1000$^\circ$/s (Lewen et al.  in prep.) and responds robustly and
reliably to angular velocities of over 2000$^\circ$/s.  
 
The arguments
presented here rested chiefly on visual inspection of the spike trains,
and this  has obvious limitations. Our eyes are drawn to reliable
features in the response, and one may  object that these cases could be
accurate but rare, so that the bulk or average behaviour of the  spike
train is much sloppier. To proceed we must turn to a more quantitative
approach. 

\section{How do we analyse the responses \\to natural stimuli?}

When we deliver simple sensory stimuli it is relatively
easy to analyse some measures of  neural response as a function of the
parameters that describe the stimulus. Faced with the  responses of a
neuron to the complex, dynamic signals that occur in nature--as in Fig.
1--what should we measure? How do we quantify the response and its
relation to the different  features of the stimulus? The sequence of
spikes from a motion sensitive neuron constitutes  an encoding of the
trajectory $\nu (t)$. Of course, this encoding is not perfect: there is
noise in the  spatiotemporal pattern of the photon flux from which motion
is computed, the visual system  has limited spatial and temporal
resolution, and inevitably there is internal noise in any  physical or
physiological system. This may cause identical stimuli to generate
different  responses. The code also may be ambiguous in the sense that,
even if noise were absent, the  same response can be induced by very
different stimuli. Conceptually, there are two very  different questions
we can ask about the structure of this code. First, we can ask about the 
features of the spike train that are relevant for the code: Is the timing
of individual spikes  important, or does it suffice to count spikes in
relatively large windows of time? Are  particular temporal patterns of
spikes especially significant? Second, if we can identify the  relevant
features of the spike train then we can ask about the mapping between
these features  of the response and the structure of the stimulus: What
aspects of the stimulus influence the  probability of a spike? How can we
(or the fly) decode the spike train to estimate the stimulus  trajectory,
and how precisely can this be done? 

There are two general approaches to
these problems. One is to compute correlation functions.  A classic
example is the method of ``reverse correlation'' in which we correlate the
spike train  with the time varying input signal (see Section 2.1 in Rieke
et al. 1997). This is equivalent to  computing the average stimulus
trajectory in the neighbourhood of a spike. Other possibilities  include
correlating spike trains with themselves or with the spike trains of
other neurons. A  more subtle possibility is to correlate spike trains
that occur on different presentations of the  same time dependent signal,
or the related idea of computing the coherence among responses  on
different presentations (Haag and Borst 1997). All of these methods have
the advantage  that simple correlation functions can be estimated
reliably even from relatively small data  sets. On the other hand, there
are an infinite number of possible correlation functions that one  could
compute, and by looking only at the simpler ones we may miss important
structures in  the data. 

An alternative to computing correlation
functions is to take an explicitly probabilistic point of  view. As an
example, rather than computing the average stimulus trajectory in the 
neighbourhood of a spike, as in reverse correlation, we can try to
characterize the whole  distribution of stimuli in the neighbourhood of a
spike (de Ruyter van Steveninck and Bialek  1988). Similarly, rather than
computing correlations among spike trains in different  presentations of
the same stimulus, we can try to characterize the whole distribution of
spike  sequences that occur across multiple presentations (de Ruyter van
Steveninck et al. 1997,  Strong et al. 1998). The probability
distributions themselves can be difficult to visualize, and  we often
want to reduce these rather complex objects to a few sensible numbers,
but we must  be sure to do this in a way that does not introduce
unwarranted assumptions about what is or  is not important in the
stimulus and response. Shannon (1948) showed that there is a unique  way
of doing this, and this is to use the entropy or information associated
with the probability  distributions. Even if we compute correlation
functions, it is useful to translate these  correlation functions into
bounds on the entropy or information, as is done in the stimulus 
reconstruction method (Bialek et al. 1991, Rieke et al. 1997, Haag and
Borst 1997, Borst and  Theunissen 1999). Although the idea of using
information theory to discuss the neural code  dates back nearly to the
inception of the theory (MacKay and McCulloch 1952), it is only in  the
last ten years that we have seen these mathematical tools used widely for
the  characterization of real neurons, as opposed to models. 

\subsection{Correlation functions}

Although we believe that the best
approach to analyzing the neural response to natural stimuli  is grounded
in information theory, we follow Warzecha and Egelhaaf and begin by using 
correlation functions. From an experiment analogous to the one in our
Fig. 2, Warzecha et al.  (1998) compute the correlation function of the
spike trains of simultaneously recorded H1 and  H2 cells, $\Phi_{\rm spike
H1-spike H2}(\tau )$, and also the average crosscorrelation function among
spike trains  \hfill from 
different \hfill presentations \hfill (trials) \hfill of the same stimulus
\hfill trajectory,  \\$\Phi_{\rm crosstrial H1-H2}(\tau )$. If the 
spike trains were reproduced perfectly from trial to trial, these two
correlation functions  would be identical; of course this is not the
case. Warzecha and Egelhaaf conclude from the  difference between the two
correlation functions that the spikes are not ``precisely time  coupled''
to the stimulus, and they argue further that the scale which
characterizes the  precision (or imprecision) of spike timing can be
determined from the width of the crosstrial  correlation function
$\Phi_{\rm crosstrial H1-H2}(\tau ) $. This is one of their arguments in
support of the notion  that the time resolution of the spike train under
natural conditions is in the range of 40 to 100  milliseconds, one or two
orders of magnitude less precision than was found in previous work. 

The crosstrial correlation function obviously contains information about
the precision of the  neural response, but there is no necessary
mathematical relation between the temporal  precision and the width of
the correlation function. To make the discussion concrete, we show  in
Fig. 3a the autocorrelation $\Phi_{\rm spike-spike}(\tau )$ and in Fig. 3b
the crosstrial correlation function  $\Phi_{\rm crosstrial}(\tau )$
computed for the outdoor experiment. We see that $\Phi_{\rm
crosstrial}(\tau )$ is very broad, while  $\Phi_{\rm spike-spike}(\tau )$
has structure on much shorter time scales, as found also by Warzecha and 
Egelhaaf. But the characterization of the crosstrial correlation function
as broad does not  capture all of its structure: rather than having a
smooth peak at $\tau = 0$, there seems to be a rather  sharp change of
slope or cusp, and again this is seen in the data presented by Warzecha
and  Egelhaaf, even though the stimulus conditions are very different.
This cusp is a hint that the  width of the correlation function is hiding
structure on much finer time scales. 

Before analyzing the correlation
functions further, we note some connections to earlier work.  Intuitively
it might seem that by correlating the responses from different trials we
are probing  the reproducibility of spike timing in some detail. But
because $\Phi_{\rm crosstrial}(\tau )$ is an average over  pairs of
spikes (one from each trial), this function is not sensitive to
reproducible patterns of  spikes such as those we have seen in Fig. 2. In
fact, the crosstrial correlation function is equal  (with suitable
normalization) to the autocorrelation function of the time dependent rate
$r(t)$  that we obtain by averaging the spike train across trials. Thus
the crosstrial correlation does  not contain information beyond the usual
poststimulus time histogram or PSTH, and the time  scales in the
correlation function just measure how rapidly the firing rate can be
modulated;  again, there is no sensitivity to spike timing beyond the
rate, and hence no sensitivity to spike  patterns. Since the crosstrial
correlation function is equal to the autocorrelation of the rate, the 
Fourier transform of  $\Phi_{\rm crosstrial}(\tau )$ is equal to the power
spectrum of the rate, which has been used  by Bair and Koch (1996) to
discuss the reproducibility of responses in the motion sensitive  neurons
of monkey visual cortex. If we Fourier transform both the crosstrial
correlation  function and the spike-spike correlation, their ratio is
proportional to the crosstrial coherence  considered by Haag and Borst
(1997) in their analysis of H1. 

Even granting the limitations of the correlation function as a probe of
spike timing, we would  like to reveal the finer time scale structure that
seems to be hiding near $\tau =0$. To do this we  consider a simple model
that can be generalized without changing the basic conclusions.  Imagine
that each spike has an ``ideal'' time $\langle t_{\rm i}\rangle$ relative
to the stimulus, and that from trial to  trial the actual arrival time of
the ith spike fluctuates as $t_{\rm i} = \langle t_{\rm i}\rangle
+\delta t_{\rm i}$. The meandering of spikes  from trial to trial in Fig.
2c suggests that the $\delta t_{\rm i}$ and $\delta t_{\rm j}$ of nearby
spikes i and j are correlated,  and if these correlations extend over a
sufficiently long time (roughly 10 ms is sufficient) then  there is a
simple approximate equation relating the crosscorrelation among trials to
the  autocorrelation and the distribution of time jitter, $P(\delta
t_{\rm i})$:
$$
\Phi_{\rm crosstrial}(\tau ) = \Phi_{\rm PP}(\tau ) \otimes
\Phi_{\rm spike-spike}(\tau ),
$$
where $\otimes$ denotes
convolution and $\Phi_{\rm PP}(\tau )$ is the autocorrelation of the
distribution $P(\delta
t_{\rm i})$. Thus  $\Phi_{\rm PP}(\tau )$ can be computed from the
measured correlation functions by deconvolution. For our  outdoor
experiment we find that $\Phi_{\rm PP}(\tau )$ has a width of 3.1 ms (Fig.
3c), so that a reasonable  estimate for the width of the underlying
jitter distribution is $\delta t_{\rm rms} = 3.1/\sqrt{2} \approx
2.2\,{\rm  ms}$. This  analysis shows that the difference between the
crosstrial and the spike-spike correlation  functions is consistent with
jitter in the range of a few milliseconds, not the many tens of 
milliseconds claimed by Warzecha and Egelhaaf.\footnote{There are further difficulties in the  interpretation of
correlation functions offered by Warzecha and Egelhaaf.  One of their
arguments for the irrelevance of high frequency stimuli is based on a
comparison of the velocity  spectrum with the spectrum of fluctuations in
the time dependent rate (Fig. 2 of Warzecha and Egelhaaf 1997);  the
spectrum of the time dependent rate should be the Fourier transform of
the crosstrial correlation function, as  noted above. On the down going
slope, across a decade of frequency the decline in the response spectrum
is  slower than the decline in the stimulus spectrum. If we define a
transfer function by taking the ratio of the  response and stimulus
spectra, then the cell is amplifying the higher frequency components, not
attentuating  them as Warzecha and Egelhaaf claim. This is consistent
with the experiments of Haag and Borst (1998)  demonstrating that the
motion sensitive neurons have active membrane mechanisms to achieve such 
amplification.}

Because the interpretation of correlation functions is a crucial issue,
let us give an example  from spatial vision, where it is clear that the
width of the correlation function (correlation  length) is not a good
indicator of the precision required to read out a signal. It is well 
documented that natural scenes typically have broad spatial correlations,
often associated with  1/f-like power density spectra (Srinivasan et al.
1982, Field 1987, Ruderman and Bialek  1994). Using the same reasoning
that Warzecha and Egelhaaf apply to spike trains, one would  conclude
that the visual system should not bother to use high spatial resolution.
This would be  true for environments with Gaussian statistics, where
second order descriptions--the simplest  correlation 
functions--are sufficient. But the world we live in 
definitely is not Gaussian. It is made out of objects 
that typically have well defined edges, and these edges 
are important to  us, not least because they are often 
associated with rigid objects. The width of the spatial 
correlation function is defined, very roughly, by the 
apparent size of the objects in our visual field. But 
this width has nothing to do with the precision with 
which we can estimate the  position of edges and hence 
the location of object boundaries. Just as for spatial 
edges, the  location of temporal edges may also be 
important, and we can look at horse racing for an  
example: In Warzecha and Egelhaaf's interpretation we 
would not need to time horses any  more precisely than 
the width of the ``horse density'' correlation function,
which corresponds  roughly to the time required for the 
entire horse to cross the finish line. Yet fortunes are 
won  and lost over differences corresponding to a 
fraction of a horse's nose. What matters here is  that 
we attach importance to features that are defined very 
sharply in time, and this temporal  precision cannot be 
measured from the width of one simple correlation 
function. For precisely  the same reason one cannot 
equate the relevant time scale of retinal image motion 
to spike  timing precision, as Warzecha and Egelhaaf 
argue in this volume (Sect. 6). Let us then turn to  an 
information theoretic approach.

\subsection{Information}

Looking at the responses to repeated presentations
of a natural complex dynamic stimulus, as  in Fig. 2, we see many
different features, some of which have been noted above: there are 
individual spikes which are reproduced from trial to trial with
considerable accuracy; there are  patterns of spikes in which the
intervals between spikes are reproduced more accurately than  the
absolute spike times, so that the patterns appear to `meander' from trial
to trial; there are  trials in which spikes are deleted, apparently at
random, and trials in which extra spikes  appear. How are we to make
sense out of this variety of phenomena? Specifically, we want to  know
whether the detailed timing of spikes is important for the encoding of
naturalistic  stimuli. How can we analyse data of this sort to give us a
direct answer to this question about  the structure of the neural code?

Intuitively, the sequence of action potentials generated by H1 ``provides
information'' about  the motion trajectory. If the response of H1 were
always the same, independent of the  trajectory, of course no information
would be provided. Generally, then, the greater the range  of possible
responses the greater is the capacity of the cell to provide information:
if we think  of segments of the neural response as being like words in a
language, then the ability of the  neuron to `describe' the input is
enhanced if it has a larger vocabulary. On the other hand, it  clearly is
not useful to generate words at random, no matter how large our
vocabulary, and so  there must be a reproducible relationship between the
choice of words and the form of the  motion trajectory. These intuitive
ideas have a precise formulation in Shannon's information  theory
(Shannon 1948): the size of the neuron's `vocabulary' is
measured by the entropy of  the distribution of 
responses, the (ir)reproducibility of the relation 
between stimulus and  response is related to the 
conditional or noise entropy computed from the 
distribution of  responses seen in multiple trials, and 
the information that the response conveys about the 
stimulus is the difference between the entropy and the 
noise entropy (see also de Ruyter van  Steveninck et al.
1997). These measures from information theory are not 
just one of many  possible ways of quantifying the 
neural response; Shannon proved that these are the only 
measures of variability, reproducibility and information
that are consistent with certain simple  and intuitively
plausible constraints.

If we believe that
the neural code makes use of a time resolution $\Delta t$, then we can
describe the  neural response in discrete time bins of this size. If
$\Delta t$ is very large this amounts to counting  the number of spikes
in each bin, while as $\Delta t$ becomes small this description becomes a
binary  string in which we record the presence or absence of individual
spikes in each bin. As our  time resolution improves (smaller $\Delta t$)
the size of the response `vocabulary' increases because  we are
distinguishing as different responses that were, at larger $\Delta t$,
lumped together as being  the same. Quantitatively, the entropy of the
responses is a function of time resolution, so that  the capacity of the
neuron to convey information is greater at smaller $\Delta t$, as first
emphasized  by MacKay and McCulloch (1952). The question of whether spike
timing is important to the  neural code is then whether neurons make
efficient use of this extra capacity (Rieke et al.  1993, 1997). In the
next section we address precisely this question in the context of the 
`outdoor' experiment on H1, reaching conclusions that parallel closely
those from our earlier  work (de Ruyter van Steveninck et al. 1997,
Strong et al. 1998). First we consider the results  of Egelhaaf and
coworkers, who have drawn nearly opposite conclusions. 

Warzecha and Egelhaaf (1997) and Egelhaaf and Warzecha
(1999) set out to study the  dependence of information transmission on
time resolution, along the lines indicated above.  Specifically, they
count spikes in bins of size $\Delta t$ and then ask how much information
this  spike count on a single trial provides about the local firing rate,
or ``Stimulus Induced  Response Component'' (SIRC) computed as an average
over many trials. Their information  measure shows a peak for a window
width of $\Delta t=80\,{\rm  ms}$ (Warzecha and Egelhaaf 1997, Fig.  3),
from which they conclude that this is the time resolution at which
signals are best  represented by H1. It is not clear what measure of
information Warzecha and Egelhaaf (1997)  are using to find the optimum:
the rate at which the spike train provides information about the 
stimulus must be a monotonic function of the time resolution. By marking
spike arrival times  more accurately we can only gain, and never lose,
information. Thus a proper measure of  information rate vs. time
resolution cannot show the behavior reported by Warzecha and  Egelhaaf.  

In the present volume they
substitute the information theoretical analysis by one in which they 
quantify the same difference (that is, between the SIRC and a running
window average count  of the single trial spike train) by a standard
deviation. This standard deviation reaches a  minimum for a window width
$\Delta t$ of about 50ms. This analysis, as their correlation function 
analysis, is based on a consideration of second order statistics, and is
therefore subject to the  same shortcomings discussed before.  

Both these
approaches suffer from the same fundamental problem: Warz\-echa and
Egelhaaf do  not quantify the relation between the neural response and
the stimulus, but instead between  spike counts and the SIRC. Implicitly,
then, they postulate that the stimulus is encoded  exclusively in the
time dependent firing rate, or the SIRC as they prefer to call it, and
further  that all information about the local rate can be ``read out'' by
counting spikes.\footnote{Even if the changing stimulus serves  only to modulate the spike
rate, it might be that different rates can be  distinguished more easily
because, for example, the shape of the interspike interval distributon
changes as  function of rate. This is known to occur in many cells.
Mathematically, counting spikes is the optimal way of  recovering rate
information only if the spike train is a modulated Poisson process. }
  As in the  analysis of crosstrial correlation
functions, this ignores by construction the possibility that  temporal
patterns of spikes may play a special role in the code, and their
reasoning is  therefore circular. For many investigators this issue of
whether patterns are important is the  question about the structure of
the neural code, and in the case of H1 it is now more than a  decade
since de Ruyter van Steveninck and Bialek (1988) reported that patterns
with short  interspike intervals carry a considerable excess of
information about the stimulus (see also  Rieke et al. 1997 and Brenner
et al. in press). The approach taken by de Ruyter van  Steveninck et al.
(1997) and by Strong et al. (1998) describes the neural response at fine
time  resolution as a binary string, marking the presence or absence of
spikes in each small time  bin, and hence all patterns of spikes are
included automatically. This is the approach that we  will use below for
the analysis of the outdoor experiment. 

In principle, the methods used by
de Ruyter van Steveninck et al. (1997) and by Strong et al.  (1998) are
independent of any model for the structure of the neural code: we do not
need to  assume that we know which features of the neural response are
relevant, nor do we need to  assume which features of the stimulus are
most important for the neuron. A number of results  on information
transmission by H1 have been obtained with a less direct method, in which
we  use the spike train to reconstruct the stimulus and then measure the
mutual information  between the stimulus and the reconstruction (Bialek
et al. 1991, Haag and Borst 1997, 1998).  Warzecha and Egelhaaf emphasize
that errors in the reconstruction result only in part from  noise, and
they claim that one therefore cannot conclude anything about the
reliability of  neurons from the quality of reconstructions (Warzecha and
Egelhaaf 1997; see also their  contribution to this volume). The thrust
of their argument is that there need be no conflict  between their claim
of imprecision in the coding of behaviourally relevant stimuli and 
previous work demonstrating precise reconstruction of the velocity
waveforms, because the  reconstruction doesn't really measure the
precision of the neural system. But this discussion  ignores the fact
that the reconstruction method provides a lower bound on the performance
of  the neuron (Rieke et al. 1997, Borst and Theunissen 1999). Thus it is
possible that  reconstruction experiments underestimate the precision of
neural coding and computation, but  properly done the reconstruction
method cannot overestimate neural performance.  

Since the reconstruction
procedure is a bound on performance and not a direct measurement,  it is
reasonable to ask how tight this bound will be. Warzecha and Egelhaaf
state that the  reconstruction of velocity signals would underestimate
the performance of the neuron if the  cell is sensitive to derivatives of
the velocity; specifically they claim that the coherence  between the
stimulus and reconstruction would be reduced if the neuron were sensitive
to  derivatives. In fact, the particular reconstruction procedure of
Bialek et al. (1991) is invariant  to linear transformations of the
signal such as differentiation and integration, and the  computation of
coherence always is invariant to these transformations (Lighthill, 1958).
Is  there any independent way to assess the efficacy of the
reconstruction method? One approach  is to try different reconstruction
algorithms (Warland et al. 1997). Another is to check for  consistency
among different measures of coherence (Haag and Borst 1997).  Finally, we
can compare the noise levels in the reconstructions with the noise levels
that  would be generated by an ideal observer who is limited only by
noise in the photoreceptors.  In the high frequency limit (of order 30
Hz), where the ideal observer's performance can be  
calculated from photoreceptor measurements, the ideal 
observer does not perform substantially better than the 
reconstruction (Bialek et al. 1991), which demonstrates 
that H1's  response approaches ideal observer 
performance. This can only be true if the fly's visual 
brain  makes efficient use of the information present in
the array of photoreceptors, and does not add a 
substantial amount of noise to the computation of 
motion. This finding is confirmed by  measurements of 
neural performance that do not depend on reconstructions
(de Ruyter van  Steveninck and Bialek, 1995). Of course,
the accuracy and efficiency of the  reconstructions  also imply the
functional correctness of the reconstruction algorithm. Criticism of the 
reconstruction algorithm itself cannot invalidate the demonstration of
accurate  reconstructions.

\section{Information transmission with natural stimuli}

In the following we use methods described in detail by  Strong et al.
(1998) to quantify  information transmission in our natural motion
experiment. Briefly, we analyse the statistics  of firing patterns that
H1 produces in response to the stimulus used in our experiment, and 
consider segments of the spike train with length $T$ divided in a number
of bins of width $\Delta t$,  where $\Delta t$ will range from very small
(order of a millisecond) up to $\Delta t=T$. Each such bin may  hold a
number of spikes, and within a bin no distinction is made on where the
spikes appear.  However, two windows of length $T$ that have different
combinations of filled bins are  considered to be different firing
patterns, and are therefore distinct. From an experiment in  which we
repeat a reasonably long natural stimulus a number of times (here 200
repetitions of  a 5 seconds long sequence) we get a large number of these
firing patterns, and from that set  we compute two entropies:  
\begin{enumerate}
\item The total entropy, which
characterizes the probability distribution of all spike firing  patterns
of length $T$ that consist of $n$ adjacent bins each $\Delta t$ wide
(that is, $T=n\Delta t$). This  entropy measures the richness of the
`vocabulary' used by H1 under these experimental  
conditions, hence the time of occurrence of the pattern 
within the experiment is irrelevant. \item The noise 
entropy, which gives us an estimate of how variable the 
response to identical stimuli can be. 
We first accumulate, for each point in time in the
stimulus sequence, the  distribution across all trials of firing patterns
that begin at that point. The entropy of this  distribution measures the
(ir)reproducibility of the response at each instant. Calculating  this
for each point in time and averaging all these values we obtain the
average noise  entropy. 
\end{enumerate}
The information contained in firing patterns of
length $T$ and resolution $\Delta t$ is the total entropy  minus the
average noise entropy (Shannon 1948). One interesting measure is to
estimate this  information as we let $T$ become very long, and $\Delta t$
very short. This limit is the average rate of  information transmission,
as discussed by Strong et al. (1998). Here, instead, we will just 
calculate the information transmitted in constant time windows,
$T=30\,{\rm ms}$, as a function of $\Delta t$.  We choose $T=30\,{\rm
ms}$ because that amounts to the delay time with which a chasing fly
follows  turns of a leading fly during a chase (Land and Collett, 1974);
the end result, namely the  dependence of information transmission on
$\Delta t$, does not depend critically on the choice of $T$.  

The data in Fig.
4a show that the information contained in a 30 ms window depends
strongly  on $\Delta t$, increasing from about 2 bits to about 5 bits
when the resolution increases from  $\Delta t =  30\,{\rm ms}$ to $\Delta
t = 1\,{\rm ms}$. Although in the limit of arbitrarily fine time
resolution ($\Delta t \rightarrow 0$), the  information must reach a
finite limit, we see no evidence for a plateau at $\Delta
t = 1\,{\rm ms}$. For shorter 
time windows ($T=12\,{\rm ms}$) we find that the information keeps on
increasing up to $\Delta t = 0.25\,{\rm ms}$.  This lack of a clear
plateau makes sense: the motion stimuli themselves have a distribution
of  temporal features so it is not surprising that there is not a sharply
defined single timescale in  the response. We also note that, as in
earlier work with less natural stimuli (Strong et al.  1998), the
information rate is a bit more than half the total entropy, even at
millisecond  resolution (see Fig. 4b), so the neuron utilizes a
significant fraction of its coding capacity  even on this fine time
scale. 

The question of whether spike timing is important
in the neural code has been debated for  decades, and our present
experiment addresses the importance of millisecond resolution in 
information transmission by a single cell. Ultimately one would like to
connect the responses  of neurons to animal behaviour. Thus, one way to
demonstrate the importance of spike timing  would be to search for
experimental conditions in which the timing of just a few spikes would 
be correlated with a behavioral decision, in the spirit of the work by
Newsome and colleagues  (Newsome et al. 1995). Another approach is to
look for other neurons that can ``read'' the  temporal structure, for
example along the lines of recent work from Usrey et al. (1998). Here  we
focus on the response of a single neuron, and ask if the precise timing
of spikes carries  information under natural stimulus conditions. The
answer is yes. 

\section{Responses to static and dynamic stimuli}

The measured
precision of responses in H1 to dynamic stimuli seems to suggest that
the  behavior of the fly visual system might be very different from other
systems, especially the  mammalian cortex. Neurons in visual cortex, for
example, commonly show a large variance  in the responses across repeated
presentations of the same visual stimulus (Tolhurst et al.  1983). To
quantify this observation several groups have studied the variance in the
number of  spikes that are counted in a window of fixed size, and then
manipulated the stimulus  conditions to find the relation between the
variance of the response and its mean. Typically,  the variance in spike
count is found to be close to or somewhat larger than the mean over a 
wide range of conditions; there is a tendency for the ratio variance/mean
(the Fano factor) to  be larger in larger time windows. 

More recently
several groups are investigating to what extent accurate spike timing,
such as  observed in H1, can be consistent with the variability of neural
responses observed in cortex.  Almost all experiments on the variability
of responses in visual cortex had been done with  static or slowly
varying stimuli, while all the work indicating precise responses and the 
importance of spike timing in H1 had been done using complex, dynamic
inputs. Newsome  and collaborators studied the responses of motion
sensitive neurons in the monkey visual  cortical area MT using dynamic
random dot stimuli, but their work focused on the connection  of neural
responses to the monkeys perception of coherent motion in the entire
display  (Newsome et al. 1995). Bair and Koch (1996) reanalysed some of
these data to show that  when the monkey saw exactly the same dynamic dot
movies the neural response showed  significant modulations on a time
scale of 30 ms or less. Strong analysed the same data to  show that the
spike train of a cortical neuron could provide information about the
movie at a  rate of $\sim 2$ bits/spike, comparable to the results in H1
(see note 19 in Strong et al. 1998), and in  unpublished analyses he found
that the variance of the spike count in windows of 30 ms or  less could
be significantly less than the mean. 

Mainen and Sejnowski (1995) found
that they could produce irregular spike trains in a slice of  cortex if
they injected constant current into a neuron: after some time the cell
`forgets' the  time at which the current was turned on and the spikes
drift relative to the stimulus. With  dynamic currents, however, there
can be precise temporal locking of spikes to particular  events in the
input signal. Berry et al. (1997) found that ganglion cells in the
vertebrate  retina---which are known to generate irregular and highly
variable spike trains in response to  static or slowly varying
images---generate highly reproducible spike trains in response to  more
dynamic movies. A hint in the same direction had been found earlier by
Miller and  Mark (1992), who showed that primary auditory neurons in the
cat give less variable  responses to complex speech stimuli than to pure
tones. Finally, de Ruyter van Steveninck et  al. (1997) showed explicitly
that the low variance, reproducible response of H1 to dynamic  stimuli
coexists with a much more variable response to constant velocity inputs:
studying a  range of constant velocities that drove H1 to average firing
rates up to about 70 spikes per  second (which corresponds to the time
average rate elicited by dynamic stimuli in comparable  stimulus
conditions), mean counts and variances in 100 ms windows straddled the
line at  which variance is equal to mean, and fell well within a cloud of
points obtained from  experiments in visual cortex. 

Taken together, all
of these different results point to the conclusion that the statistical 
structure of the neural response to static stimuli may be very different
from that in response to  dynamic or naturalistic stimuli. The crucial
conclusion is that we cannot extrapolate from the  observation of highly
variable responses under one set of conditions to reach conclusions 
about the structure of the neural code under more natural conditions.
This fits very well with  the ethological perspective that we introduced
at the beginning of this contribution, and  indeed many of the analysis
methods that we have discussed here were developed to meet the 
challenges of quantifying the neural response to more naturalistic
stimuli. From a more  mechanistic point of view there is now considerable
interest in understanding why neurons  seem to respond so differently to
static and dynamic inputs (Schneidman et al. 1999, Jensen  1998). 

Against this background it came as a surprise when Warzecha and Egelhaaf
(1999) claimed  that the variance of H1's response to constant velocity
is no different from that in response to  dynamic stimuli. It would
appear that they have done an experiment very similar to that  described
by de Ruyter van Steveninck et al. (1997) but reached the opposite
conclusion:  while Warzecha and Egelhaaf confirm the highly reproducible,
low variance response to  dynamic stimuli, they find similarly
reproducible responses to constant velocities. There are  many issues
here, but we focus first on the explicit disagreement regarding the
variability of  responses to constant velocity. Warzecha and Egelhaaf
themselves offer several possible  explanations for the discrepancy, but
they do not draw attention to the fact that the stimuli  used in the two
sets of experiments differ substantially; these differences exist along
every  stimulus dimension known to affect the response of H1---velocity,
image contrast, spatial  pattern, and size of the visual field. Further,
in the crucial comparison of static to dynamic  stimuli, it is not clear
what is being held constant in the Warzecha and Egelhaaf experiments.  In
the dynamic experiments changes in spike rate are of course driven by
variations in angular  velocity, but in their static experiments they
hold the velocity fixed and vary the image size.  At best these
experiments show that H1 responds with different statistics under
different  conditions, but we still find the discrepancies disturbing. 

In an attempt to resolve the issue, we have gone back over several years
of experiments to  collect all the data which may be relevant to relation
between variance and mean in static  experiments, we have done new
experiments that come close to the conditions of the  Warzecha and
Egelhaaf work, and we have designed new stimuli that highlight the 
differences between static and dynamic responses in a single experiment.
In brief, studying 20  flies under a wide variety of static stimulus
conditions, we find a broad distribution of  variances at each value of
the mean, but up to rates of about 100 spikes/s there is no overlap  with
the results of Warzecha and Egelhaaf (1999). Further, when we match the
conditions of  their experiments we cannot reproduce even the mean spike
counts, let alone the variances.  For example, their Fig. 4A in this
volume shows a mean spike count of about 6.5 in 100 ms  windows for a
high contrast large field ($91^\circ \times 7.5^\circ$) pattern moving at
about 36$^\circ$/s, and for the  same experimental conditions Warzecha and
Egelhaaf (1999) report a mean count of about 4.  These values correspond
to mean firing rates of 65 and 40 spikes/s respectively. In 8 flies 
tested under comparable conditions (contrast, velocity and stimulated
area) we never get rates  below 120 spikes/s, consistent with the
findings of Lenting et al. (1984).  

In Fig. 5 we show the response of H1 to a slowly varying
velocity ramp, and contrast this  response to that obtained with dynamic
velocities. Computing the mean spike counts in 100  ms windows across 50
trials, we see that the static and dynamic stimuli give the same range
of  mean responses, yet when we compute the variances there are huge
differences that are  obvious to the eye. The count variance during
quasistatic stimulation peaks for mean counts  that are about equal to
the average count during dynamic stimulation (that is, in the two places 
where the dashed line in Fig. 5b intersects the smooth curve). This is
not just a coincidence of  our choice of standard deviation of the
dynamic stimulus; it turns out that the fly's visual  system adapts such
that the mean firing rate during dynamic stimulation is rather
insensitive  to the standard deviation of the dynamic stimulus (Brenner
et al., submitted). For higher  values of the mean count the variance
decreases strongly, due to the effects of refractoriness  (Hagiwara,
1954). So H1 has relatively low count variance both for low and high
rates, but its  count variance is high for intermediate rates. Loosely,
one may think of the dynamic stimulus  as switching the cell rapidly back
and forth from a state of low rate and low variance to a state  of high
rate and low variance. By switching fast, the cell effectively bypasses
the intermediate  condition of high variance, so that its count variance
for dynamic stimuli always remains low,  as can be seen directly from
Fig. 5c. Thus, if we match windows with the same mean count,  up to a
count of about 10 for 100 ms windows, we find that H1's count variance is
lower in  response to dynamic than to static stimuli, which was precisely
the point of the original work  by de Ruyter van Steveninck et al.
(1997). 

\section{Conclusion}

Most of what we know about the nervous system has
been learned in experiments that do not  even approach the natural
conditions under which animals normally operate. Much of our  recent
work, and the core of the debate between our groups and Warzecha and
Egelhaaf,  concerns the structure of the neural code under natural
conditions. We emphasize that this is  not an easy problem, and by no
means are the issues specific to flies or even the visual  system; in
many different sensory and motor systems we would like to design and
analyse  experiments on the coding and processing of more natural
signals. 

Our approach has been to break this large question into
(hopefully) manageable pieces, and  then to use information theory as a
framework to pose these questions in a form such that  suitable
experiments should yield precise quantitative answers. In particular, we
endeavour to  make statements that do not depend on multiple prior
assumptions, and to develop methods  which can be used in analyzing many
different kinds of experiments. Thus, we have used  stimulus
reconstruction techniques to give lower bounds on the performance of fly
motion  sensitive neurons, and we have been able to measure the average
information carried by single  spikes, patterns of spikes, and continuous
segments of the spike train, all without assumptions  regarding the
``important'' features of the stimulus or neural response. Many of the
results  obtained in this way point clearly toward a picture of the fly's
visual system as close to  optimal in extracting motion information from
the photoreceptor cell array, and then encoding  this signal efficiently
in the timing of action potentials of motion sensitive neurons. In
contrast to the view developed over the past decade, the recent papers
from Egelhaaf and  coworkers, including their contribution to this
volume, make the explicit claim that the system  is very noisy and that
meaningful information is contained only in averages over time  windows
containing many spikes. In many cases these claims are introduced with
plausible  qualitative arguments. As emphasized long ago by Bullock
(1970), however, the challenge is  to quantify the degree of noisiness or
precision in the nervous system, and there is a danger  that a neuron may
appear noisy because we have an incomplete understanding of its
function.  Thus we have grown skeptical about qualitative or even
semiquantitative arguments for the  imprecision of neural responses. The
interpretation of correlation functions, discussed above,  provides a
good example: Although there may be an obvious ``correlation time'' in
one  correlation function, the hint that other time scales are relevant is
hidden in the cusp of the  correlation function at short times. More
detailed analysis shows that the relations among  correlation functions
are consistent with temporal precision on scales a factor of 30 smaller 
than the nominal correlation time. Further, this measure of temporal
precision is consistent  with the results of a rigorous information
theoretic analysis. 

Because this paper is intended (by the editors) as a
response to the contribution of Warzecha  and Egelhaaf, we have tried to
understand how they have reached conclusions so nearly  opposite from our
own. As emphasized at the outset, there are two different questions.
First  there is the problem of constructing an approximately natural
stimulus, and then there is the  problem of analyzing the response to
such a complex signal. Although there is a whole  generation of
quantitative observations on insect flight trajectories, Warzecha and
Egelhaaf  present as the stimulus they analyse a signal that is
substantially impoverished both in  amplitude and in frequency content,
as is clear from Fig. 1. Further, their visual stimulus is   very dim
compared to daytime natural conditions, and has a visible area much
smaller than  what is experienced by a free flying animal. They
repeatedly stress that motion induced  responses depend on many stimulus
variables in addition to velocity, but never discuss  critically the
extrapolation from their experiments to natural behaviour, in spite of
the large  differences in many of the crucial variables. Similarly,
although there is now a decade of  papers concerning the quantitative
information theoretic analyses of neural spike trains, and  of the cell
H1 in particular, Warzecha and Egelhaaf do not present their results on
information  transmission in absolute units (bits); closer examination
suggests that there are more basic  mathematical problems in their
approach, as outlined above.  

In this paper we have presented the results
from a new experiment (Lewen et al. in  preparation) which brings us much
closer to the natural conditions of fly vision. Visual  inspection of the
responses of H1 under these conditions indicates that individual spikes
are  reproducible on a millisecond time scale, and aspects of temporal
pattern in the spike train can  be reproducible on a substantially
submillisecond time scale. This impression is borne out by  the
quantitative demonstration that the spike train conveys information with
nearly constant  efficiency down to millisecond time resolution; indeed,
the information provided by the spike  train shows no sign of saturation
as we approach millisecond resolution. These responses to  natural
stimuli thus are even more precise than suggested by our earlier work.

In the early 1960's Reichardt and his coworkers started working on flies,
with a special  emphasis on motion detection (Reichardt, 1961). One of
their motivations was that motion  detection in flies represents a good
compromise between a reasonable complexity of  information processing
properties and an amenability to quantitative analysis (Reichardt and 
Poggio 1976). Over the years this intuition has proved to be very
fruitful, and the fly has  turned out to be a system in which many issues
could be studied, often with unprecedented  quantitative detail. In
particular, the fly's motion sensitive neurons have been an important 
testing ground for ideas about the neural code, especially in the ongoing
effort to characterize  the coding of more natural stimuli. Approached
with proper mathematical tools, the fly visual  system can continue to
provide answers to many fundamental and quantitative questions in  real
time neural information processing.

\section*{References}

\begin{description}
\item
Bialek W, Rieke F, de Ruyter van Steveninck RR, Warland, D (1991) 
Reading a neural code. {\em Science} 252:1854--1857
\item 
Bair W, Koch C (1996) Temporal precision  of spike trains in extrastriate
cortex of the behaving Macaque  monkey. {\em Neural Comp}  8:1185--1202
\item
Berry MJ, Warland DK, Meister M (1997)  The structure and precision of
retinal spike trains. {\em Proc Natl Acad Sci  USA} 94:5411--5416
\item
Borst A, Egelhaaf M (1987) Temporal modulation  of luminance adapts time
constant of fly movement detectors.  {\em Biol Cybern} 56:209--215 
\item
Borst A, Theunissen, FE (1999) Information theory and  neural coding.
{\em Nature Neurosci} 2:947--957 
\item
Brenner N, Strong SP, Koberle R, Bialek W, de
Ruyter van Steveninck R Synergy in a neural code. {\em Neural  Comp,} in
press.
 \item
Brenner N, Bialek W, de Ruyter van Steveninck R  Adaptive rescaling
maximizes information transmission.  Submitted. 
\item
Bullock TH (1970) The reliability of neurons. {\em J Gen Physiol}
55:565--584 
\item
Cover T, Thomas J (1991) {\em Elements of Information Theory.} Wiley,
New York
\item
Dethier VG (1976) {\em The hungry fly. A physiological  study of the
behavior associated with feeding.} Harvard  University Press, Cambridge
MA 
\item
Egelhaaf M, Warzecha A-K (1999) Encoding of motion  in real time by the
fly visual system. {\em Curr Opinion in  Neurobiol} 9:454--460
\item
Field D (1987) Relations between the statistics  of natural images and
the response properties of cortical cells. {\em J  Opt Soc Am A}
4:2379--2394 
\item
Gauck V, Borst A (1999) Spatial response properties of  contralateral
inhibited lobula plate tangential cells in the  fly visual system. {\em J
Comp Neurol} 406:51--71  
\item
Haag J, Borst A (1997) Encoding of visual motion
information and reliability in spiking and graded potential  neurons.
{\em J Neurosci} 17:4809--4819  
 \item
Haag J, Borst A (1998) Active membrane
characteristics and signal encoding in graded potential neurons. {\em J 
Neurosci} 18:7972-7986  
\item
Hagiwara S (1954)  Analysis of interval
fluctuations of the sensory nerve impulse. {\em Jpn J Physiol} 4,
234--240. 
\item
van
Hateren JH, Schilstra C (1999) Blowfly flight and optic flow II. Head
movements during flight. {\em J Exp Biol}  202:1491--1500 
\item
Hecht S, Shlaer S,
Pirenne MH(1942) Energy, quanta and vision. {\em J Gen Physiol}
25:819--840 
\item
Jensen, RV (1998) Synchronization of randomly driven nonlinear
oscillators. {\em Phys. Rev. E} 58:6907--6910 
\item
Land MF (1981) Optics and vision
in invertebrates. In: Autrum H (ed) {\em Handbook of Sensory Physiology 
VIII/6b.} Springer, Berlin, Heidelberg, New York, pp 472--592  
\item
Land MF,
Collett TS (1974) Chasing behavior of houseflies (Fannia canicularis). A
description and analysis. {\em J  Comp Physiol} 89:331--357  
\item
Lenting, BPM,
Mastebroek HAK, and Zaagman WH (1984) Saturation in a wide--field,
directionally selective  movement detection system in fly vision.
{\em Vision Res} 24:1341--1347 
\item
Lewen GD, Bialek W, de Ruyter van Steveninck RR Neural
coding of natural stimulus ensembles. In  preparation.  
\item
Lighthill, MJ
(1958) {\em An introduction to Fourier analysis and generalised
functions.} Cambridge University Press,  Cambridge, UK  
 \item
MacKay D, McCulloch WS (1952)
The limiting information capacity of a neuronal link. {\em Bull Math
Biophys}  14:127--135  
\item
Maddess T, Laughlin SB (1985) Adaptation of the
motion-sensitive neuron H1 is generated locally and governed  by contrast
frequency. {\em Proc R Soc Lond B} 225:251--275  
\item
Mainen ZF, Sejnowski TJ (1995)
Reliability of spike timing in neocortical neurons. {\em Science} 268:
1503--1506 
\item
Miller MI, Mark KE (1992). A statistical study of cochlear
nerve discharge patterns in response to complex  speech stimuli, {\em J
Acoust Soc Am} 92:202-209. 
\item
Newsome WT, Shadlen MN, Zohary E, Britten KH, Movshon
JA (1995) Visual motion: Linking neuronal  activity to psychophysical
performance. In: Gazzaniga M (ed) {\em The Cognitive Neurosciences.} MIT
Press,  Cambridge MA pp 401--414 
\item
Reichardt, W (1961) Autocorrelation, a
principle for the evaluation of sensory information by the central
nervous  system. In: Rosenblith WA (ed){\em  Principles of sensory
communication.} Wiley, New York, NY, pp 303--317  
\item
Reichardt W, Poggio T
(1976) Visual control of orientation behavior in the fly. Part I: A
quantitative analysis. {\em Q  Rev Biophys} 9:311--375  
 \item
Rieke F, Warland D,
Bialek W (1993) Coding efficiency and information rates in sensory
neurons. {\em Europhys  Lett} 22:151--156  
\item
Rieke F, Warland D, de Ruyter van
Steveninck RR, Bialek W (1997) {\em Spikes: Exploring the neural code.}
MIT  Press, Cambridge, MA  
 \item
Ruderman DL, Bialek W (1994) Statistics of natural
images: Scaling in the woods. {\em Phys Rev Lett} 73:814--817  
\item
de Ruyter van
Steveninck RR, Zaagman WH, Mastebroek HAK (1986) Adaptation of transient
responses of a  movement sensitive neuron in the visual system of the
blowfly Calliphora erythrocephala. {\em Biol Cybern}  54:223--236  
 \item
de Ruyter van
Steveninck R, Bialek W (1988) Real-time performance of a movement
sensitive neuron in the  blowfly visual system.{\em  Proc R Soc Lond B}
234:379--414  
 \item
de Ruyter van Steveninck R, Bialek W (1995) Reliability and
statistical efficiency of a blowfly movement--sensitive neuron. {\em Phil
Trans R Soc Lond B} 348:321--340  
\item
de Ruyter van Steveninck RR, Bialek W,
Potters M, Carlson RH, Lewen GD (1996) Adaptive movement  computation by
the blowfly visual system. In: Waltz DL (ed) {\em Natural and Artificial
Parallel Computation:  Proceedings of the Fifth NEC Research Symposium.}
SIAM, Philadelphia, pp 21--41  
\item
de Ruyter van Steveninck RR, Lewen GD,
Strong SP, Koberle R, Bialek W (1997) Reproducibility and  variability in
neural spike trains. {\em Science} 275:1805--1808  
\item
Schilstra C, van Hateren JH
(1998) Stabilizing gaze in flying blowflies. {\em Nature} 395:654  
\item
Schneidman
E, Freedman B, Segev I (1998) Ion channel stochasticity may be critical
in determining the  reliability and precision of spike timing. {\em Neural
Comp} 10, 1679--1703  
\item
Shannon CE (1948) A mathematical theory of
communication. {\em Bell Syst Techn J} 27:379--423 and 623--656 
\item
Srinivasan MV,
Laughlin SB, Dubs A (1982) Predictive coding: A fresh view of inhibition
in the retina. {\em Proc R  Soc Lond B} 216:427--459  
\item
Strong SP, Koberle R, de
Ruyter van Steveninck RR, Bialek W (1998) Entropy and information in
neural spike  trains. {\em Phys Rev Lett} 80:197--200 
\item
Tolhurst DJ, Movshon JA, Dean AF (1983) The statistical reliability of signals in single neurons in cat and 
monkey visual cortex. {\em Vision Res} 23:775--785.
\item
Usrey WM, Reppas JB, Reid RC (1998)  Paired-spike interactions and
synaptic efficacy of retinal inputs to the  thalamus. {\em Nature}
395:384--387 
\item
Wagner H (1986a) Flight performance and visual control of flight of the free-flying housefly (Musca domestica 
L.). I. Organization of the flight motor. {\em Phil Trans R Soc Lond B}
312:527--551 
\item
Wagner H (1986b) Flight performance and visual control of flight of the
free-flying housefly (Musca domestica  L.). II Pursuit of targets. {\em
Phil Trans R Soc Lond B} 312:553--579  
\item
Wagner H (1986c) Flight performance and
visual control of flight of the free-flying housefly (Musca domestica 
L.). III. Interactions between angular movement induced by wide- and
smallfield stimuli. {\em Phil Trans R Soc Lond B} 312:581--595  
 \item
Warland DK,
Reinagel P, Meister M (1997) Decoding visual information from a
population of retinal ganglion  cells. {\em J Neurophysiol} 78:2336--2350 
\item
Warzecha A-K, Egelhaaf M (1997) How reliably does a neuron in the visual
motion pathway of the fly encode  behaviourally relevant information?
{\em Europ J Neurosci} 9:1365--1374  
\item
Warzecha A-K, Egelhaaf M (1998) On the
performance of biological movement detectors and ideal velocity  sensors
in the context of optomotor course stabilization. {\em Visual Neurosci}
15:113--122  
\item 
Warzecha A-K, Kretzberg J, Egelhaaf M (1998) Temporal
precision of the encoding of motion information by  visual interneurons.
{\em Curr Biol} 8:359--368 
\item
Warzecha A-K, Egelhaaf M (1999) Variability in spike
trains during constant and dynamic stimulation. {\em Science} 
283:1927--1930 
\item
Wehrhahn C (1979) Sex-specific differences in the chasing behavior of
houseflies (Musca). {\em Biol Cybern} 32:239--241  
\item
Wehrhahn C, Poggio T,
Bulthoff H (1982) Tracking and chasing in houseflies (Musca). An analysis
of 3-D flight  trajectories. {\em Biol Cybern} 45:123--130 
\end{description}

\vfill\newpage
\leftline{\large\bf Figure Captions}
\parindent = 0pt
\parskip = 10 pt
\bigskip

{\bf Figure 1.} Comparison of the rotational velocity  traces reported
from free flying and tethered flies. a: Rotation  velocity of a fly
(Fannia canicularis) in free flight, derived from video recordings by
Land and Collett (1974). b:  Rotation velocity of a pattern in a flight
simulator, derived from torque signals measured from a tethered fly, as 
reported by Warzecha and Egelhaaf (1997). c: The data from a and b
plotted on the same scale. 

{\bf Figure 2.} Direct observations of H1 spike timing statistics in
response to rotational motion derived from Land  and Collett's (1974)
free flight data (see Fig. 1a). a: A 500 ms segment of the motion trace.
b: Top: raster plot  with 25 traces representing spike occurrences
measured from H1. Bottom: raster plot of 25 traces of spike  occurrences
from the same cell, but in response to a velocity trace that was the
negative of the one shown in a.  For ease of reference we call these
traces H1$+$ and H1$-$ respectively. c: Raster plot of 25 samples of the 
occurrence time of the first spike fired by H1$-$ after time $t=2080$ ms
in the stimulus sequence (indicated by the  dashed line connecting the
axis of b to panel c). d: Raster plots of 25 samples of the interval from
the spike  shown in c to the first (filled circles), second (open
circles), third (filled triangles), and fourth (open triangles)  spike
following the spike shown in c. Note the time axes: The rasters in c and
d are plotted at much higher time  resolution than those in b. e:
Probability density for the timing of the spike shown in c. The spread is 
characterized by $\sigma = 0.73$ ms, where $\sigma$ is defined as half the
width of the peak containing the central 68.3\% of  the total probability.
If the distribution were Gaussian, then this would be equivalent to the
standard deviation.  Here we prefer this definition instead of one based
on computing second moments. The motivation is that there  can be an
occasional extra spike, or a skipped spike, giving a large outlier which
has a disproportionate effect on  the width if it is calculated from the
second moment. Filled squares represent the experimental histogram,
based  on 200 observations, while the solid line is a Gaussian fit. f:
Probability densities for the same interspike interval  shown in d. The
definition of $\sigma$ is the same as the one in e.

{\bf Figure 3.} Correlation functions for H1 during stimulation with
natural motion, all computed at 0.2 ms resolution.  a: The spike--spike
autocorrelation $\Phi_{\rm spike-spike}(t)$, normalized as a conditional
rate. There are strong oscillations in  the conditional rate, due to
neural refractoriness. b: The cross-trial correlation function
$\Phi_{\rm crosstrial}(t)$, computed as  the correlation function of the
estimated time dependent rate minus a contribution from $\Phi_{\rm
spike-spike}(t)$ scaled by  $1/N$ ($N$ is the number of trials) to correct
for intratrial correlations. c: Autocorrelation of the assumed
underlying  distribution of spike jitter times, computed by deconvolving
the data in b by those in a. See text for further  explanation.

{\bf Figure 4.} Information and coding efficiency in firing patterns for
naturalistic motion stimuli (see legend for Fig.  2). a: Total entropy,
noise entropy and information in an observation window of $T=30\,{\rm
ms}$, as a function of time  resolution, $\Delta t$. From the trends
observed in partitioning the finite dataset we estimate first and second
order  extrapolations to the entropies for an infinite dataset. Filled
symbols are first order, open symbols are second  order extrapolations.
The deviation between first and second order extrapolations is small,
indicating that  systematic errors in the entropy estimates are small
(for details see Strong et al. 1998). Statistical errors were  estimated
from the spread in the different partitions of the original dataset.
These errors are smaller than the size  of the symbols. b: Coding
efficieny  (information divided by total entropy) as a function of
$\Delta t$.

{\bf Figure 5.} Mean count and variance compared for quasistatic and
dynamic velocity stimuli. a: Stimulus. for the  first 48 s the velocity
is slowly ramped up and down. From 50 to 72 seconds the stimulus is
dynamic with a  standard deviation of  100$^\circ$/s, and a cutoff
frequency of 250 Hz. Note that the vertical scales (left for the 
quasitationary  and right for the dynamic stimulus) are different. The
peak at 50 s is a reset phase in which the  pattern is moved at maximum
speed so as to bring the stimulus pattern into exact register on every
trial. b: Trial  average spike count in 100 ms windows, as a function of
time. The dashed line represents the time averaged  count in response to
the dynamic stimulus. c: Spike count variance in 100 ms windows, as a
function of time. 

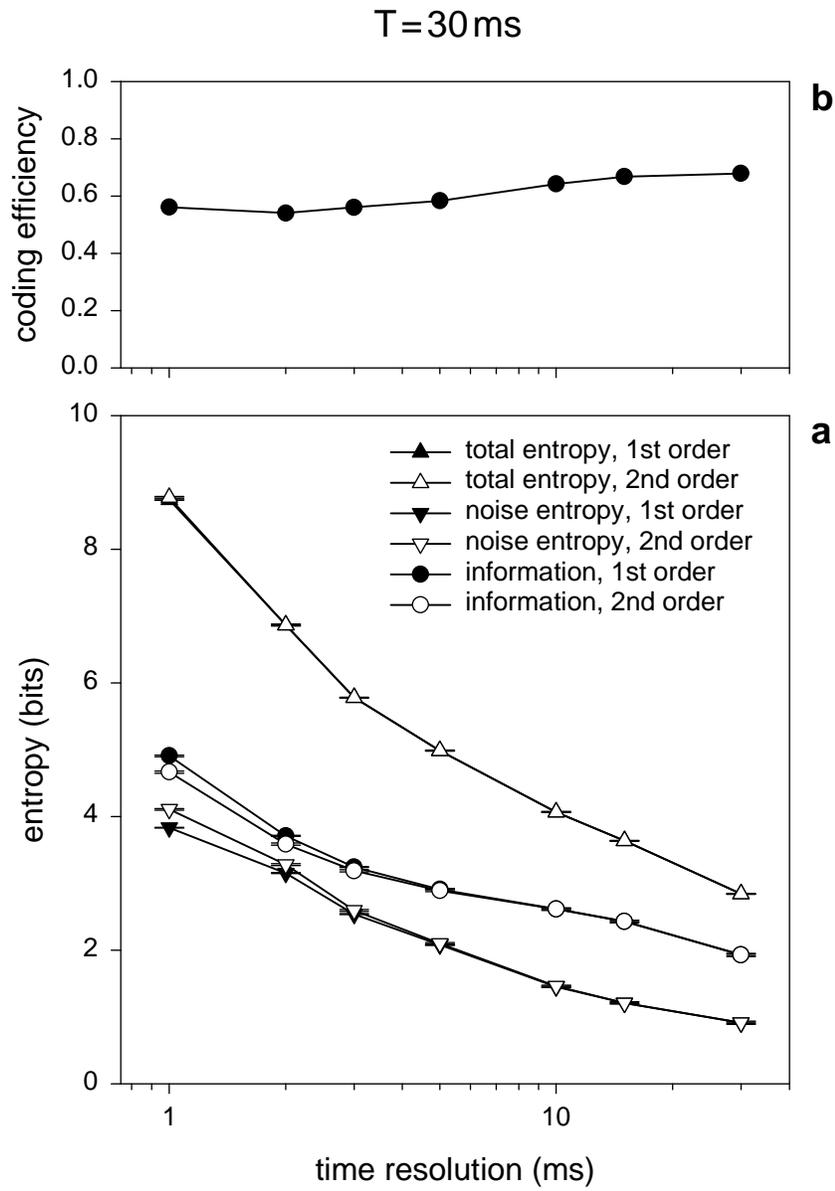
\end{document}